\documentstyle[11pt,aaspp4]{article} 

\def\deg{\ifmmode^\circ\else$^\circ$\fi}

\def\kpc{\ifmmode h^{-1}{\rm kpc}\else$h^{-1}{\rm kpc}$\fi}
\def\kms{\ifmmode {\rm km~s}^{-1}\else${\rm km~s}^{-1}$\fi}
\def\feh{\ifmmode {\rm [Fe/H]}\else [Fe/H]\fi}

\lefthead{Chiba}
\righthead{Structure of the Galactic Stellar Halo}

\begin{document}

\title{Structure of the Galactic Stellar Halo Prior to Disk Formation}

\author{Masashi Chiba}
\affil{National Astronomical Observatory, Mitaka, Tokyo 181-8588, Japan\\
email: chibams@gala.mtk.nao.ac.jp}
\author{Timothy C. Beers}
\affil{Department of Physics \& Astronomy, Michigan State University, E.
Lansing, MI 48824\\email:  beers@pa.msu.edu}

\begin{abstract}
We develop a method for recovering the global density distribution of the
ancient Galactic stellar halo prior to disk formation, based on the {\it
present} orbits of metal-poor stars observed in the solar neighborhood.
The method relies on the adiabatic invariance of the action integrals of motion
for the halo population during the slow accumulation of a disk component,
subsequent to earlier halo formation. The method is then applied to a sample of
local stars with [Fe/H]$\le-1.5$, likely to be dominated by the halo component,
taken from Beers et al.'s recently revised and supplemented catalog of
metal-poor stars selected without kinematic bias.  We find that even if the
Galactic potential is made spherical by removing the disk component in an
adiabatic manner, the halo density distribution in the inner halo region ($R
\le 15$ kpc) remains moderately flattened, with axial ratio of about 0.8
for stars in the abundance range [Fe/H]$\le -1.8$ and about 0.7 for the more
metal-rich interval $-1.8<$[Fe/H]$\le-1.5$.  The outer halo remains spherical
for both abundance intervals.  We also find that this initial flattening of the
inner halo is caused by the anisotropic velocity dispersions of the halo stars.
These results suggest that the two-component nature of the present-day stellar
halo, characterized by a highly flattened inner halo and nearly spherical outer
halo, is a consequence of {\it both} an initially two-component density
distribution of the halo (perhaps a signature of dissipative halo formation)
{\it and} of the adiabatic flattening of the inner part by later disk
formation.  Further implications of our results for the formation of the Galaxy
are also discussed, in particular in the context of the hierarchical clustering
scenario of galaxy formation.

\end{abstract}


\section{Introduction}

The currently favored cold dark matter (CDM) theory of galaxy formation
postulates that the formation of a massive spiral galaxy like our own is a
consequence of the hierarchical assembly of subgalactic dark halos, and the
subsequent accretion of cooled baryonic gas in a virialized, galaxy-scale dark
halo (e.g., Peacock 1999). Numerical studies based on this picture are able to,
at least qualitatively, reproduce the characteristic features of a disk galaxy
-- the massive dark halo, the stellar halo, and the stellar disk components
(e.g., Steinmetz \& M\"uller 1995; Bekki \& Chiba 2000; Navarro \& Steinmetz
2000), though difficulties are still encountered in the details.  For example,
the simulations conducted to date do not adequately account for the size of the
disk component and the number of satellite galaxies (Navarro, Frenk, \& White
1995; Moore et al. 1999; Klypin et al. 1999).

The CDM hierarchical model may be regarded, in its essence, as a modern
generalization of the classical Searle \& Zinn (1978) hypothesis for the
formation of the Galactic stellar halo.  To explain a large inferred spread in
the ages of globular clusters, and the lack of a spatial gradient in their
metal abundances, Searle \& Zinn argued that the halo component may have
experienced prolonged, chaotic accretion of subgalactic fragments, as opposed
to the rapid, monolithic collapse proposed by Eggen, Lynden-Bell, \& Sandage
(1962). Recent discoveries of halo substructures in Galactic phase space
(Majewski, Munn, \& Hawley 1994; 1996; Helmi et al. 1999; Chiba \& Beers 2000,
hereafter CB; Yanny et al. 2000) and of the Sagittarius dwarf galaxy, which is
presently being disrupted by the Galactic tidal field (Ibata, Gilmore, \& Irwin
1994; Ibata et al. 2000), may lend further support to this picture.

Although halo formation via hierarchical assembly of subgalactic systems, such
as dwarf galaxies, may continue to the present day (Bland-Hawthorn \& Freeman
2000), a large fraction of the stellar halo, especially the inner part
where the disk lies, should have been completed {\it prior to disk
formation}, since otherwise the disk component is made significantly thicker
than is observed due to dynamical heating from infalling masses (Toth \&
Ostriker 1992). A clear age gap between the (thin) disk and the stellar halo
supports that the latter consists of ancient populations (e.g. Liu \& Chaboyer
2000). Also, studies of star-forming histories in the disk component
indicate that the disk has been accumulated at an approximately constant rate
over the last several billion years (Twarog 1980; Sommer-Larsen \& Yoshii
1990); frequent mergings of dwarf galaxies over the Galaxy's lifetime will
entirely modify the photometric and spectroscopic properties of the disk.
Thus, one may well postulate that the inner part of the stellar halo, at say $R
\le 15$ kpc, retains a fossil imprint of how it was formed.

The question then arises, what {\it was} the structure of the halo component
prior to disk formation?  Since the bulk of halo stars are found in this inner
part, where the disk gravity is dominant, the present-day structure of the halo
can be greatly affected by later disk formation.  It is thus necessary to
consider the dynamical effect of the disk in inferring the structure of the
ancient halo from the currently observed halo stars.

Binney \& May (1986, hereafter BM) examined this issue by assuming adiabatic
invariance (for halo stars) of the action integrals of motion ${\bf J}$,
and for the distribution function $f({\bf J})$ during slow disk formation.
They set up test particles distributed in a spheroid, similarly to the stellar
distributions of elliptical galaxies, and calculated the dynamical response of
the particles to the slow increase of disk mass inside the spheroid.  They
showed that the Galactic halo, before the disk was formed, may have had a
somewhat flattened density distribution (axial ratio $q \sim 0.7$), in order to
produce a current highly flattened halo ($q \sim 0.3$), which they inferred
from the radially anisotropic velocity ellipsoid of local metal-poor stars.

We note here that recent kinematic data for larger samples of local metal-poor
stars indicate a more moderately flattened halo ($q \sim 0.7$) for inner radii
($R \le 15$ kpc), whereas the outer halo is nearly spherical (Sommer-Larsen \&
Zhen 1990, hereafter SLZ; CB). This is also supported by examination of the
spatial distributions of other halo tracers (e.g., Hartwick 1987; Preston et
al.  1991; Kinman, Suntzeff, \& Kraft 1994; Yanny et al. 2000).  Also, the
extent to which the formation of the disk component flattened the halo depends
on the unknown initial velocity distribution of halo stars, whereas the initial
conditions set up by BM apply only for one specific case.  Thus, it is yet
unexplored what the currently available data for metal-poor stars may tell us
about the structure of the ancient halo before the disk was formed.

In this paper we revisit this issue, based on a large sample of halo
stars in the solar neighborhood, taken from a recently completed catalog
of metal-poor stars selected without kinematic bias (Beers et al. 2000).
It is noted that in a similar vein, Sommer-Larsen (1986), in his thesis work,
arrived at a conclusion similar to BM's by investigating the distribution of
individual orbital inclinations for his sample of 143 stars with
[Fe/H]$\le-1.2$. In contrast, we seek herein to develop a more general and
direct method, based on the BM picture, to calculate the global density
distribution of the halo prior to disk formation. The method is then applied
to the more accurate and numerous data for metal-poor stars that is
presently available.

This paper is organized as follows. In \S 2 we describe general properties of
orbits in the mass model of the St\"ackel type that we adopt here, as well as
the methodology for constructing the global density of a given local sample
both before and after disk formation. The mass model for the Galactic
potential that we adopt, and the local sample of metal-poor stars used in the
current analysis, are also described.  In \S 3 we compute the orbital motions
of the sample stars while conserving action integrals of motion.  We then
present the adiabatic change of the derived density distributions and kinematic
properties of the sample when the disk component is slowly removed (essentially
working backwards from the present to the past).  Finally, in \S 5, the results
are summarized, and their implications for the formation and evolution of the
Galaxy are discussed.

\section{Method and Sample}

In this work, we adopt the axisymmetric Galactic potential of the St\"ackel
type, for which the Hamilton-Jacobi equation separates in spheroidal
coordinates (e.g., de Zeeuw 1985; Dejonghe \& de Zeeuw 1988).  Although this
type of potential omits the resonant orbits and accompanied
stochastic orbits that can be
revealed in a more general non-separable potential, the fraction of such
orbits, which are basically associated with the 1:1 resonance in the radial and
vertical directions, is thought to be small in the Galactic phase space (May \&
Binney 1986; BM). Moreover, in contrast to a non-separable potential, for which
extensive numerical integrations of orbits are required, the analytic nature of
the St\"ackel type model has the great advantage of maintaining clarity in the
analysis.  The role of resonant and stochastic orbits revealed from the sample
we investigate below will be discussed elsewhere (Allen et al. 2000).

\subsection{St\"ackel Models and Orbital Densities}

In the following, we briefly describe the basic properties of the St\"ackel
models and refer the reader to, e.g., de Zeeuw (1985) and Dejonghe \& de Zeeuw
(1988), for more details.

We construct the axisymmetric Galactic potential of the St\"ackel type,
which is defined in spheroidal coordinates $(\lambda,\phi,\nu)$, where $\phi$
corresponds to the azimuthal angle in cylindrical coordinates
$(R,\phi,z)$,
and $\lambda$ and $\nu$ are the roots for $\tau$ of
\begin{equation}
\frac{R^2}{\tau+\alpha} + \frac{z^2}{\tau+\gamma} = 1 \ ,
\end{equation}
where $\alpha$ and $\gamma$ are constants, giving
$-\gamma\le\nu\le-\alpha\le\lambda$. The coordinate surfaces are spheroids
$(\lambda=const.)$ and hyperboloids of revolution $(\nu=const.)$ with the
$z$-axis as the rotation axis, where the focal distance
$\Delta=(\gamma-\alpha)^{1/2}$ fixes the coordinate system.

The gravitational potential of the St\"ackel type is then written as
\begin{equation}
\psi(\lambda,\nu) = -
 \frac{(\lambda+\gamma)G(\lambda)-(\nu+\gamma)G(\nu)}{\lambda-\nu} \ ,
\end{equation}
where $G(\tau)$ is an arbitrary function. In this work, $G(\tau)$ is the
sum of $G_D(\tau)$ from a disk and $G_H(\tau)$ from a massive dark halo.

The Hamiltonian per unit mass, $H$, for motion in the potential
$\psi(\lambda,\nu)$ is written as
\begin{equation}
H = \frac{p_\lambda^2}{2P^2} + \frac{p_\phi^2}{2R^2}
  + \frac{p_\nu^2}{2Q^2} + \psi(\lambda,\nu) \ ,
\end{equation}
where $P$ and $Q$ are the metric coefficients of the spheroidal coordinates,
given by
\begin{equation}
P^2=\frac{\lambda-\nu}{4(\lambda+\alpha)(\lambda+\gamma)}, \
Q^2=- \frac{\lambda-\nu}{4(\nu+\alpha)(\nu+\gamma)}  \ ,
\end{equation}
and $p_\lambda$, $p_\phi$, and $p_\nu$ are the conjugate momenta to
$\lambda$, $\phi$, and $\nu$, respectively,
\begin{equation}
p_\lambda=P^2\dot{\lambda}=Pv_\lambda,  \
p_\phi=R^2\dot{\phi}=R v_\phi,\
p_\nu=Q^2\dot{\nu}=Qv_\nu \ .
\end{equation}
The velocities $v_\lambda$, $v_\phi$, and $v_\nu$ at a point
$(\lambda,\phi,\nu)$ are the components of the velocity ${\bf v}$ along
the orthogonal axis defined locally by spheroidal coordinates.

The ``standard'' three integrals of motion, ${\bf I} \equiv (E,I_2,I_3)$, are
defined as
\begin{eqnarray}
E   &=& - H \\
I_2 &=& \frac{L_z^2}{2} \\
I_3 &=& \frac{1}{2}(L_x^2+L_y^2) + \Delta^2
   \left[ \frac{1}{2}v_z^2 - z^2 \frac{G(\lambda)-G(\nu)}{\lambda-\nu}
   \right] \ .
\end{eqnarray}
Another useful set of integrals in the current study are the action integrals,
${\bf J} \equiv (J_\lambda,J_\phi,J_\nu)$, which are defined as
\begin{eqnarray}
J_\lambda &=& \frac{1}{2\pi} \oint p_\lambda d\lambda =
         \frac{2}{\pi} \int^{\lambda_2}_{\lambda_1} p_\lambda d\lambda \\
J_\phi    &=& \frac{1}{2\pi} \oint p_\phi d\phi = L_z \\
J_\nu     &=& \frac{1}{2\pi} \oint p_\nu d\nu =
         \frac{2}{\pi} \int^{\nu_0}_{-\gamma} p_\nu d\nu \ ,
\end{eqnarray}
where ($\lambda_1,\lambda_2)$ and $\nu_0$ are the turning points of the orbit,
defined as the values for which $v_\lambda=0$ and $v_\nu=0$, respectively,
and $\nu=-\gamma$ defines the equatorial plane. For the evaluation of
$J_\lambda$, we have taken four times the integrals from $\lambda_1$ to
$\lambda_2$, to maintain symmetry between $J_\lambda$ and $J_\nu$ and
ensure continuity of the actions across transitions from one orbital
family to another (de Zeeuw 1985).

With a set of three integrals of motion, {\bf I}$\equiv(E,I_2,I_3)$,
a distribution function $f({\bf x},{\bf v})$ can be expressed as
$f(E,I_2,I_3)$ by application of Jeans' theorem, provided that the system
has reached dynamical equilibrium.  Note that we expect this description 
to be valid in the inner halo of the Galaxy, where the dynamical effect of
later accreting materials is small. We employ here an ensemble of metal-poor
stars which act as tracers moving in the Galactic volume. In such a discrete
case, consisting of $N$ stars, $f$ is written as (Statler 1987),
\begin{equation}
f = \sum_{i=1}^{N} c_i \delta(E-E_i)\delta(I_2-I_{2,i})\delta(I_3-I_{3,i}) \ ,
\end{equation}
where $c_i$ is the orbit weighting factor to be determined as described below.

The density $\rho({\bf x})$ at any point ${\bf x}$ is then given as,
\begin{equation}
\rho({\bf x}) = \sum_{i=1}^{N} c_i \rho_{orb}(E_i,I_{2,i},I_{3,i}; {\bf x}) \ ,
\end{equation}
where $\rho_{orb}$ is the density of a single orbit
\begin{equation}
\rho_{orb}(E,I_2,I_3; {\bf x}) = \frac{2\sqrt{2}}{R}
     \frac{1}{ \sqrt{N(\lambda)(\lambda+\gamma)}
             \sqrt{-N(\nu)(\nu+\gamma)} \sqrt{I_2} } \ ,
\end{equation}
with
\begin{equation}
N(\tau) = G(\tau) - \frac{I_2}{\tau+\alpha}
        - \frac{I_3}{\tau+\gamma} - E \ .
\end{equation}

With the aid of $J_\tau = J_\tau(E,I_2,I_3)$, a distribution function can also
be expressed as $f(J_\lambda,J_\phi,J_\nu)$. When we define $f$ in action space
as
\begin{equation}
f = \sum_{i=1}^{N} c_i^\ast \delta(J_\lambda-J_{\lambda,i})
\delta(J_\phi-J_{\phi,i}) \delta(J_\nu-J_{\nu,i}) \ ,
\end{equation}
where $c_i^\ast$ is the orbit weighting factor in action space,
the density $\rho$ is now given as
\begin{equation}
\rho({\bf x}) = \sum_{i=1}^{N} c_i^\ast \rho_{orb}^\ast
(J_\lambda,J_\phi,J_\nu; {\bf x}) \ ,
\end{equation}
where $\rho_{orb}^\ast$ is the density of a single orbit for a given
${\bf J}$. We note that $\rho_{orb}^\ast$ cannot generally be expressed in an
analytic form, as was done in eq.(14) for $\rho_{orb}$, so we will use
$\rho_{orb}$ to calculate the density distribution of the stellar halo in this
study.

Comparing eq.(13) with eq.(17) and noting
$\rho_{orb}^\ast = |\partial {\bf I}/\partial {\bf J}|
\rho_{orb}$, we obtain the relation between the amplitudes $c_i^\ast$ and $c_i$
as (Statler 1987)
\begin{equation}
c_i^\ast = c_i \left| \frac{\partial(J_\lambda,J_\phi,J_\nu)}
{\partial(E,I_2,I_3)} \right|_i
 \ .
\end{equation}

\subsection{Construction of the Global Density of the Current Stellar Halo}

The formulation given above indicates that, once the orbit weighting factor
$c_i$ is determined, we can construct the global density of the halo
for a given set of orbits, using eqs.(13)-(14). For this purpose we follow
the strategies argued by May \& Binney (1986), as implemented in the
maximum likelihood approach developed by SLZ, as explained below.

In the case of a continuous distribution function, $f({\bf J})$, the
probability at ${\bf x}$ of a star with actions ${\bf J}$, or equivalently,
the normalized density distribution of its orbit, is given by
(May \& Binney 1986)
\begin{equation}
P_{orb}({\bf x}|{\bf J}) = \frac{1}{8\pi^3}
 \left| \frac{\partial {\bf J}}{\partial {\bf v}} \right|^{-1}
  = \frac{1}{8\pi^3} \rho_{orb}^\ast \ .
\end{equation}
Then, from Bayes' theorem, the probability that a star found at ${\bf x}$
has actions in the range $\delta^3{\bf J}$ centered on ${\bf J}$ is given
by
\begin{equation}
dP = \frac{ P_{orb}({\bf x}|{\bf J})f({\bf J})\delta^3{\bf J} }
     { \int P_{orb}({\bf x}|{\bf J})f({\bf J})\delta^3{\bf J} } \ .
\end{equation}
In the discrete case, consisting of $N$ stars at positions
${\bf x}_1$,...,${\bf x}_N$, with a distribution function given in eq.(16),
the probability given above leads to (SLZ)
\begin{equation}
P_{ij} = \frac{c_i^\ast \rho_{orb,i}^\ast({\bf x}_j)}
        {\sum_{k=1}^N c_k^\ast \rho_{orb,k}^\ast({\bf x}_j)}
       = \frac{c_i \rho_{orb,i}({\bf x}_j)}
        {\sum_{k=1}^N c_k \rho_{orb,k}({\bf x}_j)} \ .
\end{equation}
Then, $c_i$ is determined by maximizing the probability that the star found
at ${\bf x}_{j=1}$ is on orbit $i=1$, the star found at ${\bf x}_{j=2}$ is
on orbit $i=2$, and so forth. The likelihood function to be maximized is
\begin{equation}
\ln \prod_{i=1}^N P_{ii} = \sum_{i=1}^{N} \ln P_{ii}
 = \sum_{i=1}^N \left[ \ln c_i \rho_{orb,i}({\bf x}_i)
   - \ln \sum_{k=1}^N c_k \rho_{orb,k}({\bf x}_i) \right] \ ,
\end{equation}
where $\rho_{orb}$ is estimated from eq.(14). Thus, given a set of
$({\bf x}_i, {\bf v}_i)$, $i=1$,...,$N$, we can calculate the integrals of
motion $(E_i,I_{2,i},I_{3,i})$ from eqs.(6)-(8), evaluate $c_i$ from eq.(22)
with $\rho_{orb}$ from eq.(14), and then obtain the global density distribution
from eq.(13).

\subsection{Method for Recovering the Ancient Halo before Disk Formation}

We now extend the above method to obtain the global density distribution of the
halo prior to disk formation, under the assumption that the Galactic disk has
been formed slowly compared with the dynamical timescale of the system, often
taken to be on the order of $10^8$ yr.  This assumption is well
supported by various lines of evidence (Twarog 1980; Sommer-Larsen \& Yoshii
1990; Rocha-Pinto et al. 2000).  In such a case, both the action
integrals ${\bf J}$ [eqs.(9)-(11)] and the distribution function $f({\bf J})$
[eq.(16)] are invariant, whereas ${\bf I}$ and $f({\bf I})$ change accordingly.

The method is summarized as follows: (1) At the current epoch,
the gravitational potential is composed of a disk and a dark halo, $G=G_D+G_H$,
and with this potential, we compute the $N$ sets of integrals
$(E_i,I_{2,i},I_{3,i})$ and  $(J_{\lambda,i},J_{\phi,i},J_{\nu,i})$,
the orbit weighting factors $c_i$, and the Jacobians
$|\partial {\bf J}/\partial {\bf I}|_i$. We then determine the orbit
weighting factors, $c_i^\ast$, which are adiabatic invariants.
(2) At the epoch prior to disk formation, the gravitational potential is
supposed to be provided by a dark halo alone, $G=G_H$. With this potential, we
search for new $N$ sets of integrals
$(E_i^\prime,I_{2,i}^\prime,I_{3,i}^\prime)$ under the condition that the
action integrals $(J_{\lambda,i},J_{\phi,i}, J_{\nu,i})$ for each star are
conserved. We then calculate the Jacobians $|\partial {\bf J}/\partial {\bf
I^\prime}|_i$ and estimate the new orbit weighting factors $c_i^\prime$ using
eq.(18).  (3) With $(E_i^\prime,I_{2,i}^\prime,I_{3,i}^\prime)$ and
$c_i^\prime$ and the dark halo potential, we obtain the global density of the
stellar halo using eq.(13).

We note that among the new sets of integrals $(E_i^\prime,I_{2,i}^\prime,
I_{3,i}^\prime)$, $I_2^\prime = I_2$ because $I_2$ is expressed as
$I_2 = J_\phi^2 / 2$ from eq.(7) and eq.(10). The search of the integrals
$(E_i^\prime,I_{3,i}^\prime)$ for a given set of the action integrals ${\bf J}$
requires numerical procedures, except for some specific forms of the potential
(see e.g., Evans, de Zeeuw, \& Lynden-Bell 1990). In our experiment using
the sample described below, the action integrals $J_\lambda$ and $J_\nu$ are
well-conserved, within a precision of $O(10^{-5})$ in their values.

\subsection{The Galaxy Model}

We now construct the Galaxy model, consisting of a disk and a dark halo, where
both components are of the St\"ackel type. Among the existing Galaxy models of
the St\"ackel type in the literature, we adopt the model originally constructed
by Dejonghe \& de Zeeuw (1988) and later elaborated by Batsleer \& Dejonghe
(1994, hereafter BD), which takes a Kuzmin-Kutuzov
potential for both a highly flattened disk and a nearly spherical halo,
\begin{equation}
G_D(\tau) = \frac{G_{grav}kM}{\sqrt{\tau}+\sqrt{-\gamma}}, \
G_H(\tau) = \frac{G_{grav}(1-k)M}{\sqrt{\tau+b}+\sqrt{-\gamma+b}}, \
\end{equation}
where $M$ is the total mass, and $k$ is the ratio of the disk mass to the
total mass. The parameter $b$ is given so as to let this two-component model
remain of the St\"ackel type.  Note that when we use $a_D$ and $c_D$, instead
of $\alpha$ and $\gamma$, to define the coordinate system for the disk
component, as $\alpha = -a_D^2$ and $\gamma = -c_D^2$, the corresponding
quantities for the halo component, $a_H$ and $c_H$, must be set as $a_H^2 =
a_D^2 + b$ and $c_H^2 = c_D^2 + b$.

To make this mass model resemble the real Galaxy, we set the following
conditions: (1) the circular velocity around the Galactic center $v_{rot}$ is
nearly constant at $\simeq 220$ km s$^{-1}$ beyond $R \simeq 4$ kpc, (2) the
local mass density at the Sun $(R_\odot,z_\odot)=(8.5,0)$ kpc is $0.1 \sim 0.2$
$M_\odot$ pc$^{-3}$ (Bahcall 1984; Bahcall et al. 1992), and (3) the surface
mass density at the solar radius is about 70 $M_\odot$ pc$^{-2}$ within $z
\le 1.1$ kpc from the plane (Kuijken \& Gilmore 1991;  Bahcall et al. 1992).
After some experimentation, the following parameter values are adopted as
a standard case: $c_D=0.052$ kpc, $c_D/a_D=0.02$, $c_H=17.5$ kpc,
$c_H/a_H=0.99$, $M=10^{12}$ $M_\odot$, and $k=0.09$. Thus the disk mass is $9
\times 10^{10}$ $M_\odot$, comprising 9 \% of the total mass.  Figure 1 shows a
rotation curve derived from the adopted parameters\footnote{These parameters
are basically
the same as those adopted by BD, except that the rotation curve in our model is
approximately flat to well beyond $R \simeq 20$ kpc, whereas the BD model
shows a falling rotation curve beyond this radius. We note that in the BD
model, some number of the sample stars we will use below are unbound to the
Galaxy because of the insufficient mass at outer radii. In our model, all
sample stars are bound to the Galaxy, although the mass distribution beyond
$R \simeq 20$ kpc is essentially irrelevant to our modeling of the stellar halo
inside this radius.},
demonstrating that the rotation curve
is nearly flat beyond $R \simeq 4$ kpc, where $v_{rot}(R_\odot) = 220$ km
s$^{-1}$. The local mass density at the Sun is 0.16 $M_\odot$ pc$^{-3}$ and the
surface mass density at $R=R_\odot$ is 68 $M_\odot$ pc$^{-2}$ within $z \le
1.1$ kpc from the plane.  The axial ratios of the potential, $q_\psi$, obtained
from these parameters, are 0.84 at $R=10$ kpc, 0.95 at $R=20$ kpc, and nearly 1
at larger radii. If we remove the disk component from the potential, $q_\psi$
is very close to 1 at all radii.

\subsection{The Sample of Local Metal-Poor Stars}

To construct the global density of the stellar halo based on the orbits of
local metal-poor stars, it is important to avoid kinematic bias in the
selection of the sample.  If the halo stars are selected based on high proper
motions, for example, the sample will have a bias against stars which show
similar orbital motions to the Sun, and the global density constructed from
such a sample will also be biased. It is similarly important to use a large and
homogeneously analyzed sample, to minimize statistical fluctuations in the
derived density distribution, and to avoid, or at least minimize,
other systematic errors.

Beers et al. (2000) presented a large catalog of metal-poor stars with
[Fe/H]$\le-0.6$, selected without kinematic bias.
A subset of 1214 stars in the catalog contain accurate proper motions taken
from recently completed proper motion catalogs, including the {\it Hipparcos}
catalog (ESA 1997), in addition to other homogeneously analyzed data with
updated stellar positions, newly-derived homogeneous distance
estimates, revised radial velocities, and refined metal-abundance estimates.
This is by far the largest sample of metal-poor field stars with available
proper motions among extant non-kinematically selected samples.
Thus, the sample is most advantageous for our current study.

We select, as representatives of the halo population, the stars in the sample
within the abundance range [Fe/H]$\le-1.5$, which is sufficiently metal-poor
to avoid contamination from stars with disk-like kinematics (Chiba \& Yoshii
1998; CB).  Also, to minimize the effects of distance errors, we confine
ourselves to the stars with measured distances $D \le 2.5$ kpc and with
rest-frame velocities $\le 550$ km s$^{-1}$, which is a likely range to bind
stars inside the Galaxy. The latter limit excludes only two stars.
Furthermore, in order to investigate whether there is a finite difference
between the density distributions of more metal-poor and metal-rich halo
populations, as argued by Sommer-Larsen (1986), we arbitrary split the sample
into two abundance ranges, [Fe/H]$\le-1.8$ and $-1.8<$[Fe/H]$\le-1.5$, as a
standard case. The effect of changing the abundance intervals on the result
will be given in the later section.
After applying these cuts, the sample we investigate includes $N=321$ stars for
[Fe/H]$\le-1.8$, and $N=182$ stars for $-1.8<$[Fe/H]$\le-1.5$.  Local
kinematics of the sample in the solar neighborhood are characterized by
radially anisotropic velocity dispersions, $(\sigma_R,\sigma_\phi,\sigma_z)=
(153\pm6,115\pm5,97\pm4)$ km s$^{-1}$ for [Fe/H]$\le-1.8$ and
$(147\pm8,114\pm6,81\pm4)$ km s$^{-1}$ for $-1.8<$[Fe/H]$\le-1.5$, and slow
systematic rotation, $<V_\phi>=31\pm6$ km s$^{-1}$ and $37\pm8$ km s$^{-1}$,
for the respective abundance ranges.

\section{Results}

In this section we investigate the adiabatic change of orbital properties and
density distributions of our sample stars when the disk mass is removed
adiabatically from the total Galactic potential.

\subsection{Adiabatic Change of the Individual Orbits}

For each star in the potential, after and before disk formation, we compute
apo- and peri-Galactic distances along the Galactic plane
($R_{ap,k},R_{pr,k}$), and the maximum height away from the plane $z_{max,k}$,
where $k=1$ for the current potential with the disk and $k=2$ when the disk is
removed adiabatically.  In Figures 2a and 2b we show the change of these
distances, demonstrating that before disk formation, the halo stars orbit at
systematically larger distances, and similarly, the spatial ranges of the
pre-disk orbits are larger than for the current epoch.  This is explained by
the fact that when a disk is slowly formed in the inner region of the
proto-Galactic sphere, the gravitational potential becomes more centrally
concentrated, and both of the integrals $(E,I_3)$ are accordingly increased, so
that the allowed regions for orbital motions are reduced and shifted toward the
Galactic center.  The characteristic expansion factor in $R_{ap}$, when the
disk is removed, is estimated as $R_{ap,2}/R_{ap,1} \simeq 1.3$. It is also
noted that the change of the potential is more pronounced in the $z$-direction
because of the disk geometry. As a result, orbital inclinations with respect to
the plane will be reduced after disk formation (Yoshii \& Saio 1979). This is
actually observed in our calculations, as demonstrated in Figure 2c, where we
plot the change of inclination angles as defined by $\zeta = \arctan
(z_{max}/R_{ap})$.

We also compute the orbital eccentricities, defined as
$e=(r_{ap}-r_{pr})/(r_{ap}+r_{pr})$, where $r_{ap}$ and $r_{pr}$ denote apo-
and peri-Galactic distances from the Galactic center, respectively, and plot
them in Figure 2d. It is apparent that the orbital eccentricities derived here
show only a little change, especially for $e<0.4$ and $e>0.8$, even though both
$r_{ap}$ and $r_{pr}$ change substantially. Thus, the orbital eccentricities,
defined arbitrarily as above, are approximately adiabatic invariants during
slow disk formation, as was also demonstrated by Eggen, Lynden-Bell, Sandage
(1962) and Yoshii \& Saio (1979).

\subsection{Global Density Distributions of the Stellar Halo}

Following the method outlined in \S 2, we now calculate the global density
distributions of the stellar halo, both at the current epoch and before disk
formation.  As was done by SLZ and CB, we proceed to average the density
distributions derived from eq.(13) over grids of finite area in the meridional
plane of the spheroidal coordinates, $(\lambda,\nu)$.  The grids are defined as
$\lambda_k=k^2-\alpha$, $k=1,...,30$ and $\nu_l=(\gamma-\alpha)
\cos^2(\theta_l) -\gamma$, $\theta_l=(\pi/2)(l/20)$, $l=0,...20$. The spatial
resolution of the grids is about 1 kpc.

In Figure 3a we plot, for the [Fe/H]$\le-1.8$ sample, the radial density
distributions along the Galactic plane (the averaged density over the area at
$l=20$) at the current epoch (open circles), and when the disk is removed
adiabatically (filled circles).  As was shown by SLZ and CB, the density
distribution for $R >8$ kpc is well described by a power-law model $\rho
\propto R^{\beta}$. At the current epoch, we find an exponent
$\beta=-3.4$ over $8 \le R \le 30$ kpc, in good agreement with the
results by SLZ and CB. Below $R=8$ kpc, the density distributions clearly
deviate from a single power-law model, a result which is likely caused by
incomplete representation of stars with apocentric radii, $R_{ap}$, below $R
\simeq R_\odot$ (SLZ; CB) in the local samples we investigate.  When the disk
is removed adiabatically, the density distributions are made shallower: we
obtain an exponent $\beta=-3.0$ over $10 \le R \le 30$ kpc, where the
lower radius for this estimate, 10 kpc, is increased from 8 kpc by taking into
account the characteristic expansion factor of $R_{ap}$ obtained in the
previous subsection. Thus, as expected, the density distribution of the stellar
halo is made more centrally concentrated when the disk is slowly formed in the
central region of the dark halo.  For $-1.8<$[Fe/H]$\le-1.5$, power-law models
with exponents $\beta=-3.3$ at the current epoch, and with
$\beta=-3.0$ when the disk is removed, provide excellent fits to the data
(Figure 3b).

In order to obtain a typical error in the estimate of the exponent $\beta$,
which arises from the combined effects of observational errors in positions and
velocities of the sample stars, we have constructed ten sets of ``pseudo-data''
for positions and velocities, where each value is randomly selected within its
standard observational error with respect to its mean value. From independent
analysis of these ten reconstructed models, we find a rms error of 0.14 in the
determination of $\beta$.

Figures 4a and 4b show the equidensity contours of the constructed global
density distributions in the $(R,z)$ plane for [Fe/H]$\le-1.8$ and
$-1.8<$[Fe/H]$\le-1.5$, respectively, at the current epoch (left panels)
and before disk formation (right panels).  The lack of stars at small $R$ and
large $z$ (which gives rise to the ill-formed contour levels in this portion of
the diagram) is again a consequence of the small probability that stars in the
Galaxy which explore such a region are represented in the solar neighborhood,
as argued in SLZ and CB.  This is also seen in Figure 2c, where the inclination
angles of orbits with respect to the plane, $\zeta$, are mostly confined to
less than about 45$^\circ$, as is the global density inferred from such orbits.
Excluding this region, these equidensity contours suggest clearly that the
current density distributions are flattened at inner radii and round at outer
radii, as was obtained by CB using a different Galactic potential.  In
contrast, when the disk is removed adiabatically, the density distributions are
made rounder, especially at inner radii.

To be more quantitative, we fit elliptical contours to the constructed density
maps while excluding the region with polar angle $\theta\le 45^\circ$.
Specifically, we obtain fits to ellipses of major axis $a$ and axial ratio $q$.
The change of our estimate of $q$ as a function of radius is shown in Figures
5a and 5b for [Fe/H]$\le-1.8$ and  $-1.8<$[Fe/H]$\le-1.5$, respectively, where
the error bars correspond to the rms errors from the best fits.

First, at the current epoch (open circles), the axial ratio $q$ remains small
at $R < 15$ kpc and increases with $R$ at larger radii, in good agreement with
CB.  For [Fe/H]$\le-1.8$, we obtain $q\simeq 0.70-0.75$ at $R < 15$ kpc and
$q\simeq 0.95$ at $R=20$ kpc.  For $-1.8<$[Fe/H]$\le-1.5$, we obtain $q\simeq
0.50-0.60$ at $R < 15$ kpc and $q\simeq 1.0$ at $R=18$ kpc. The latter
subsample appears to show a dip in $q$ at $R \simeq 20$ kpc, which we believe
is a statistical fluctuation due to the limited size of the sample ($N=182$).
It should be noted that, at $R\simeq 20-25$ kpc, the tangential anisotropy of
the velocity dispersions begins to dominate (Sommer-Larsen et al. 1997), so
that the probability that the stars which explore to large $z$ are represented
in the solar neighborhood may be small for $R > 20$ kpc. Except for such large
radii, it is interesting that the more metal-rich halo subsample shows a more
flattened density distribution at inner radii than the more metal-poor halo
subsample.  This is consistent with the somewhat smaller vertical velocity
dispersion $\sigma_z$ at $R=R_\odot$, in the former (81 km s$^{-1}$) as
compared to the latter (97 km s$^{-1}$).

Second, when the disk is removed adiabatically (filled circles), the axial
ratio is increased to $q\simeq 0.80$ for [Fe/H]$\le-1.8$ and $q\simeq 0.70$ for
$-1.8<$[Fe/H]$\le-1.5$. This result indicates that the density distribution
of the stellar halo in the inner part of the halo is rounder before disk
formation. It should be noted here that this explanation is valid in the
{\it global} sense: for instance, at $R\simeq 20$ kpc in Figure 5a, the axial
ratio is actually decreased before disk formation, which is caused by the
radial expansion of the inner, more flattened region by removing
the disk component.  The result also indicates that the axial ratio $q$ before
disk formation, in the assumed spherical potential with $q_\psi = 1$, is
smaller than $q_\psi$, possibly due to the anisotropic velocity ellipsoid of
the stars even before disk formation (see van der Marel 1991 for examples
of the comparison between $q$ and $q_\psi$). Also, there is a tendency of a
more flattened density distribution for $-1.8<$[Fe/H]$\le-1.5$ than for
[Fe/H]$\le-1.8$ even before disk formation, although its significance is too
small to be certain.

In addition to the standard case shown above, we investigate various cases
by changing the model parameters for the potential or the conditions for the
sample selection. Table 1 shows the axial ratios of the derived density
distributions after ($q_1$) and before ($q_2$) disk formation, where, for the
sake of straightforward comparison, we have calculated an average axial
ratio over $8<R<13$ kpc for $q_1$ and $8<R<17$ kpc for $q_2$. First, when
the disk mass is made smaller than the standard case ($k=0.09\to 0.08$),
the present-day potential becomes more spherical, and so $q_1$ is increased.
The amount of the change from $q_1$ to $q_2$ is the same as or
somewhat smaller than the standard case, while a small difference in $q_2$
between two abundance ranges still remains. Second, a similar change of
this result is also seen when we employ the more flattened dark halo
($c_H/a_H=0.99\to 0.98$). In this case, the contribution of the dark halo to
the midplane potential is increased at a given radius, in a manner that the
shape of the potential is made more spherical than the standard case, as would
pertain to a less massive disk.  Third, we adopt the more restrictive distance
cut ($D\le 2.5$ kpc $\to 1$ kpc), to eliminate distant stars for which the
errors in the estimated velocities are generally larger. It follows from Table 1
that while the result for [Fe/H]$\le-1.8$ remains essentially the same, both
$q_1$ and $q_2$ are systematically decreased for $-1.8<$[Fe/H]$\le-1.5$: the
difference in $q_2$ between two abundance ranges is increased. Fourth, we
change the abundance ranges so that the boundary value dividing the two ranges
is decreased ([Fe/H]$_{div}=-1.8\to -1.9$ or $-2.0$). Since the more metal-poor
stars are included in the more metal-rich subsample as [Fe/H]$_{div}$ is
decreased, the difference in the axial ratios turns out to be reduced compared
to the standard case.

To summarize, the currently flattened density distribution of
the stellar halo in the inner region is due both to adiabatic flattening
caused by the slowly formed disk and to an initially (slightly) flattened
density distribution. The tendency that the more metal-rich halo sample
exhibits a larger flattening even before disk formation may offer a clue
for understanding the formation process of the stellar halo (see \S 4).

\subsection{Adiabatic Change of the Velocity Dispersions}

We now compute the velocity dispersions of the stars at any point ${\bf x}$
predicted by the current model (following Dejonghe \& de Zeeuw 1988 and SLZ),
\begin{eqnarray}
\sigma_\tau^2 &=& \frac{1}{\rho({\bf x})}
   \sum_{i=1}^N c_i \rho_{orb}({\bf x}) v_{\tau,i}^2({\bf x}) \ ,
    \ \tau=\lambda, \nu                                      \\
\sigma_\phi^2 &=& \frac{1}{\rho({\bf x})}
   \sum_{i=1}^N c_i \rho_{orb}({\bf x})
   \left[ v_{\phi,i}({\bf x}) - \frac{1}{\rho({\bf x})}
   \sum_{j=1}^N c_j \rho_{orb}({\bf x}) v_{\phi,j}({\bf x}) \right]^2  \ ,
\end{eqnarray}
where the velocities $(v_\lambda,v_\phi,v_\nu)$ are defined by
\begin{equation}
v_\lambda= \pm \sqrt{\frac{2(I_3^+-I_3)}{\lambda-\nu}}, \
v_\phi   = \pm \frac{\sqrt{2I_2}}{R}, \
v_\nu    = \pm \sqrt{\frac{2(I_3-I_3^-)}{\lambda-\nu}} \ ,
\end{equation}
with
\begin{eqnarray}
I_3^+ &=& (\lambda+\gamma)[G(\lambda)-E]
        - \frac{\lambda+\gamma}{\lambda+\alpha} I_2  \\
I_3^- &=& (\nu+\gamma)[G(\nu)-E]
        - \frac{\nu+\gamma}{\nu+\alpha} I_2  \ .
\end{eqnarray}

Figure 6a shows the predicted velocity dispersions of the stars with
[Fe/H]$\le-1.8$, in the Galactic plane ($z=0$) at radii of $R=8.5$ and
15.5 kpc. The abscissa denotes the axial ratio of the potential, $q_\psi$,
at these radii: $q_\psi=0.80$ at $R=8.5$ kpc and 0.93 at $R=15.5$ kpc
when the disk is in place, and $q_\psi=1.00$ at both radii before disk
formation. Each line shows the change of $\sigma_\tau$
($\tau=\lambda,\phi,\nu$) after and before disk formation, with endpoints
drawn as filled circles for $R=8.5$ kpc and open circles for $R=15.5$ kpc.
For instance, at $R=8.5$ kpc, $(\sigma_\lambda,\sigma_\phi,\sigma_\nu)
=(158, 113, 112)$ km s$^{-1}$ at the current epoch (left-hand filled circles)
and $(107, 83, 64)$ km s$^{-1}$ before disk formation (right-hand filled
circles). We note here that based on the ten pseudo-data sets described in
\S 3.2, we find rms errors of these velocity dispersions as $(4, 3, 4)$
km s$^{-1}$ at $R=8.5$ kpc and $(16, 3, 4)$ km s$^{-1}$ at $R=15.5$ kpc.
Analogous to panel (a), Figure 6b shows the ratios of these velocity
dispersions $(\sigma_\phi/\sigma_\lambda,\sigma_\nu/\sigma_\lambda)$ at the
same radii.  These figures, which are to be compared with Figures 4 and 5 in
BM, indicate that the velocity dispersions increased as the disk was formed, as
a consequence of a more centrally concentrated potential arising from the disk
component.  Also, as panel (b) shows, $\sigma_\nu$ is boosted more readily than
$\sigma_\phi$ due to the flattening of the potential, so that
$\sigma_\nu/\sigma_\lambda$ increases.  This change of $\sigma_\nu$ is more
prominent at inner radii, where the disk mass density is large. On the other
hand, the ratio $\sigma_\phi/\sigma_\lambda$ decreases as the disk was formed.
This may be explained in the following way.  The epicycle theory of orbits
(e.g., Binney \& Tremaine 1987) shows that the ratio of velocity dispersions
along the plane, $\sigma_\phi/\sigma_R$, is approximately equal to
$\kappa/2\Omega$, where $\kappa$ and $\Omega$ denote epicyclic and rotational
frequencies, respectively, at a given radius.  Thus, since $\kappa/2\Omega$
decreases if the mass distribution responsible the potential is made more
centrally concentrated, $\sigma_\phi/\sigma_R$ (or equivalently
$\sigma_\phi/\sigma_\lambda$ near the plane) decreases.

Although these results are basically consistent with the general trend of the
BM model, there are notable differences in the values obtained. BM showed that
their $\sigma_\theta$ along the polar angle is only 38 km s$^{-1}$ at
$R=R_\odot$ before disk formation, suggesting significant anisotropic
flattening of the initial density distribution ($q \simeq 0.7$).  In contrast,
the corresponding value using our $\sigma_\nu$ is 64 km s$^{-1}$ and the
density is more moderately flattened ($q \simeq 0.8$). Thus, the velocity
dispersions of the halo component before the disk was formed are characterized
by an anisotropic velocity ellipsoid, but its degree of anisotropy is more
moderate than previously obtained.

The more metal-rich abundance range, $-1.8<$[Fe/H]$\le-1.5$, gives basically
the same result as the above, except for the ratios of the velocity dispersions
before disk formation --
$(\sigma_\phi/\sigma_\lambda,\sigma_\nu/\sigma_\lambda)$ at $R=8.5$ kpc is
(0.59,0.44), for $-1.8<$[Fe/H]$\le-1.5$, but takes values (0.77,0.60) for
[Fe/H]$\le-1.8$. Thus, the former, more metal-rich range, shows a more
anisotropic velocity ellipsoid than the latter. This is consistent with the
more flattened density distribution for the former, even before disk formation,
as was shown in the previous subsection.

\section{Discussion and Conclusion}

The global structure of the present-day stellar halo is characterized by an
inner, highly flattened part, as revealed at $R<15$ kpc, and an outer, nearly
spherical part (SLZ; CB). This two-component picture for the present-day
stellar halo provides a reasonable explanation why faint-star-count studies
have generally yielded an approximately spherical halo, whereas the local
anisotropic velocities of the halo stars suggest a highly flattened system
(Freeman 1987).  The issue relevant here is what physical mechanism in the
early stage of the Galaxy gives rise to the inner, highly flattened halo, where
the bulk of halo stars are found, and where the effects of later satellite
accretion may be diminished.

One of the possible reasons for the two-component nature of the present-day
halo is the slow formation of the disk within an initially spherical stellar
halo (BM). In this paper, we have quantified the effect of later disk formation
on the halo flattening, based on methods assuming adiabatic invariance of the
motion of halo stars, and its application to a large sample of stars in
the solar neighborhood. We have found that, even before disk formation, the
inner part of the stellar halo exhibited a finite flattening, although
it is more moderate than presently observed. The axial ratios of the density
profiles within the almost spherical potential are $q \simeq 0.80$ for
[Fe/H]$\le-1.8$ and $q \simeq 0.70$ for $-1.8<$[Fe/H]$\le-1.5$. Also, the
initial velocity dispersions are characterized by an anisotropic velocity
ellipsoid, as
$(\sigma_\phi/\sigma_\lambda,\sigma_\nu/\sigma_\lambda)=(0.77,0.60)$ for
[Fe/H]$\le-1.8$ and (0.59,0.44) for $-1.8<$[Fe/H]$\le-1.5$ at $R=R_\odot$.
Therefore, the inner part of the stellar halo was flattened by velocity
anisotropy, and the more metal-rich population likely exhibited a more
flattened density distribution.

Through a comparison of the inclination angles of orbits for two abundance
ranges, [Fe/H]$\le-1.5$ and $-1.5<$[Fe/H]$\le-1.2$, Sommer-Larsen (1986) also
obtained a more flattened initial distribution for more metal-rich populations.
We note that his latter subsample shows a rapid systematic rotation of 147 km
s$^{-1}$, possibly contaminated by the stars belonging to the metal-weak thick
disk (Freeman 1987). In contrast, our metal-rich halo subsample is selected
from the more restrictive range $-1.8<$[Fe/H]$\le-1.5$, where the effect of
disk-like kinematics is minimal (Chiba \& Yoshii 1998; CB) -- the flattening of
this subsample is caused by velocity anisotropy, not systematic rotation.

The results presented here may suggest that the ancient halo, at least in its
inner part, may have undergone a somewhat ordered contraction.  This
contraction could have involved dissipation due to baryonic gas -- radiative
cooling of this gas was most efficient in the innermost regions with high
density.  The resultant contraction of this gas, as a whole, may have
proceeded mainly along the axis of rotation, because of the absence of the
angular-momentum barrier in this direction.  As the chemical enrichment
proceeded along with the progress of the collapse, more metal-enriched stars
would have ``seen'' a more flattened density distribution.  On the other hand,
the outer part of the halo may have been more susceptible to later infall of
satellite galaxies, so that its density, kinematics, and mean age are different
from those of the inner halo (Norris 1994; Carney et al. 1996; Sommer-Larsen et
al. 1997).

Alternatively, one might argue that the two-component nature of the present-day
stellar halo is entirely a consequence of satellite accretion {\it after}
disk formation (Freeman 1987). According to Quinn \& Goodman (1986) (see also
Quinn, Hernquist, \& Fullagar 1993), prograde satellite orbits that are
initially inclined at less than about 60$^\circ$ to the disk are dragged down
quickly toward the plane by the effects of the dynamical friction of the disk.
Walker, Mihos, \& Hernquist (1996) further explore the effects of mergers of
small satellites with large disk galaxies such as the Milky Way.  The debris
from these merging satellites, in combination with disrupted disk stars, would
be expected to form a flattened system.  Furthermore, if more massive
satellites were more metal-rich (as they appear to be at present, see Mateo
1998), their orbits would fall farther toward the Galactic center so that their
debris would form a more flattened, more metal-rich system (Freeman 1987).
However, there exists no clear evidence for the predicted dynamical heating of
the thin-disk component -- its very thin geometry (Toth \& Ostriker 1992), and
the nearly constant velocity dispersion of thin-disk stars over the last 10
Gyrs (Quillen \& Garnett 2000), suggest that the thin disk has sustained little
significant damage since its formation.  In this regard, one might argue that
the metal-weak thick disk is evidence for early dynamical heating of a
pre-existing, metal-deficient, thin disk. However, it is then difficult to
explain its absence in the abundance range considered in this paper (Chiba \&
Yoshii 1998; CB).

It is worthwhile to remark that the hypothesis of the dissipative formation of
the inner flattened halo, as well as the later accretion of satellites onto the
outer halo, is a natural consequence of the CDM hierarchical clustering model
(Bekki \& Chiba 2000).  This model postulates that a protogalactic system
initially contains numerous subgalactic clumps, comprised of a mixture of gas
and dark matter, and that the merging of these clumps led to a smaller number
of more massive clumps.  In the simulations of Bekki \& Chiba (2000), these
larger clumps move gradually toward the central region of the system, due to
both dynamical friction and dissipative merging with smaller clumps.  Finally,
the last merging event occurs between the two most massive clumps, and the
metal-poor stars which have been formed inside the clumps are disrupted and
spread over the inner part of the halo.  The aftermath is characterized by a
flattened density distribution.  Some fraction of the disrupted gas from the
clumps may settle into the central region of the system, and produce a more
enriched, more flattened density distribution.  Some of the initially small
density fluctuations in the outer region would have gained systematically
higher angular momentum from their surroundings, and then slowly fallen into
the system after most parts of the system were formed.   This may correspond to
the process of late satellite accretion, contributing primarily to the outer
part of the halo.  Thus, the reported initial state of the stellar halo can be
explained, at least qualitatively, in the context of hierarchical clustering
scenario.

An alternative approach for elucidation of the dissipative nature of halo
formation is to examine the results of recent high-resolution N-body
simulations of structure formation based on the CDM theory (e.g., Ghigna et al.
1998; Moore et al. 1999; Klypin et al. 1999). Such simulations provide the
orbital properties of dark matter particles inside virialized dark halos.  If
the stellar halo component in the Galaxy is formed similarly through
dissipationless hierarchical assembly, the orbits of halo stars prior to disk
formation, as derived in the current paper, may follow those of dark matter
particles. For this purpose, we take the Ghigna et al. (1998) simulation of the
formation of a cluster, as this is currently the only published one that
presents the detailed orbital distribution of the simulated particles.  They
showed that the orbital distribution of the halo particles is close to
isotropic -- circular orbits are rare and radial orbits are common.  The
average ratio of pericentric and apocentric distances, $r_{pr}/r_{ap}$, is
equal to or less than 0.20, without showing a large variation as a function of
the distance from the cluster: the median ratio is approximately 0.17. On the
other hand, as is deduced from Figure 2, the orbits of the halo stars we have
derived here before the disk was in place are more circular than the simulated
dark halo particles, and the velocity field is anisotropic: the average value
of $r_{pr}/r_{ap}$ is 0.29.  Thus, we require some additional process, possibly
dissipative interaction among protogalactic clumps to circularize their orbits,
to explain the characteristic orbital distribution of the halo stars in the
early Galaxy.

More quantitative conclusions must await more elaborate modeling of the
formation of the Galaxy over a large number of possible model parameters. Also,
it is necessary to assemble and analyze the data of more remote stars,
especially those presently found inside the solar radius, where our modeling
of the halo is incomplete. In this regard, the next generation of astrometric
satellites, such as {\it FAME} and {\it GAIA}, will provide highly precise
parallaxes and proper motions for numerous stars, so that both
three dimensional positions and velocities will be available over a large
fraction of the halo.  Also, with these astrometric satellites, we will be able
to determine the exact mass distributions of the disk and dark halo components,
and thus obtain definite information on the early Galaxy, using the technique
outlined here.  Furthermore, in addition to the Milky Way, direct
identification of halo populations in external disk galaxies may prove
promising as a way to clarify the global structures of stellar halos and their
association with disks and bulges (e.g., Morrison 1999). Such studies should be
eagerly pursued with 10m class telescopes.

\acknowledgments

We are grateful to the anonymous referee for constructive comments on the
paper. This work has been supported in part by Grants-in-Aid for Scientific
Research (09640328) from the Ministry of Education, Science, Sports and
Culture of Japan.

\clearpage

\begin{figure}
\begin{center}
\leavevmode
\epsfxsize=0.8\columnwidth\epsfbox{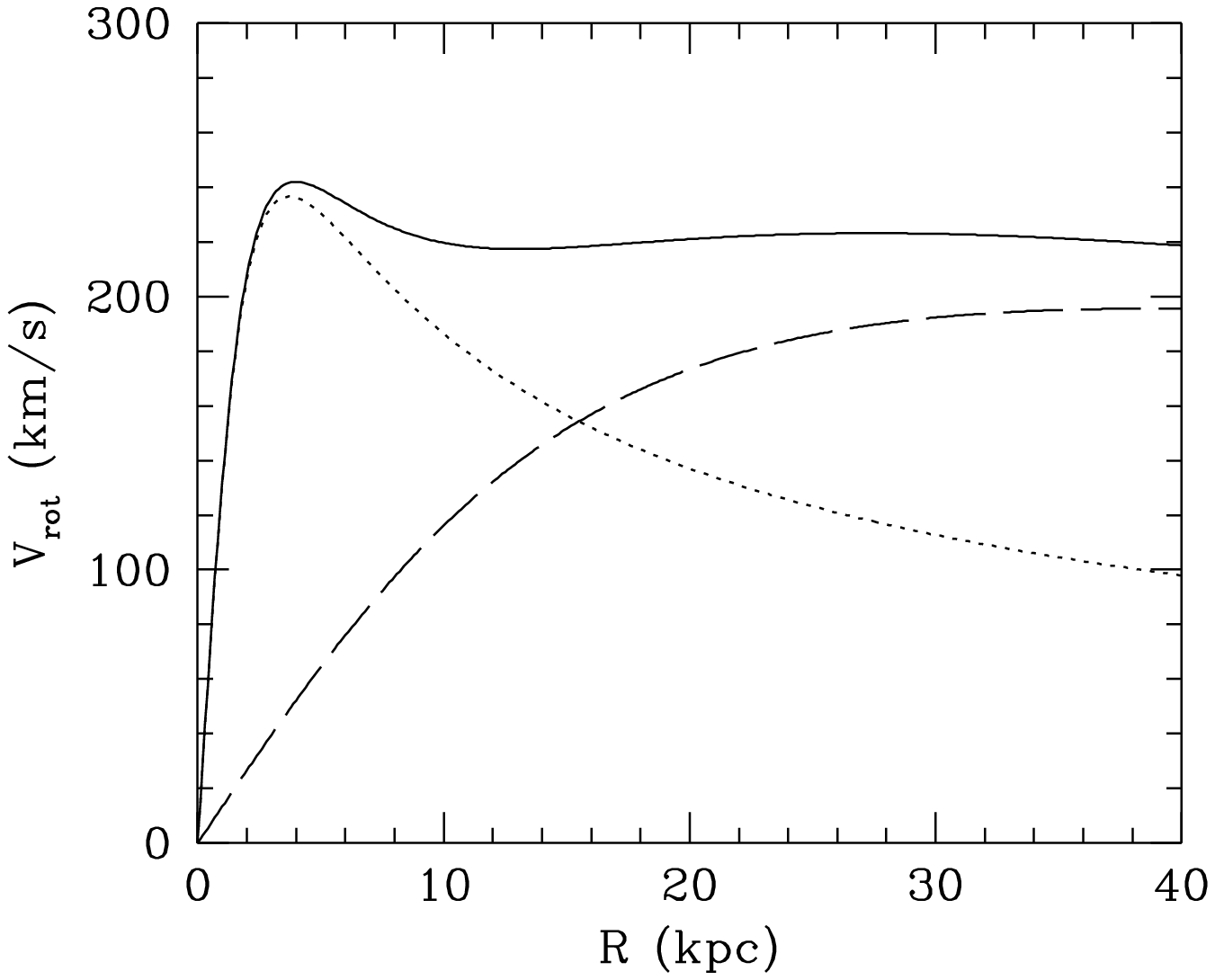}\hfil
\end{center}
\figcaption[fig1.eps]{
Rotation curve for the adopted Galactic potential
(solid line). Dotted and dashed lines denote the contributions from the
disk and dark halo, respectively.}
\end{figure}

\clearpage
\begin{figure}
\begin{center}
\leavevmode
\epsfxsize=0.8\columnwidth\epsfbox{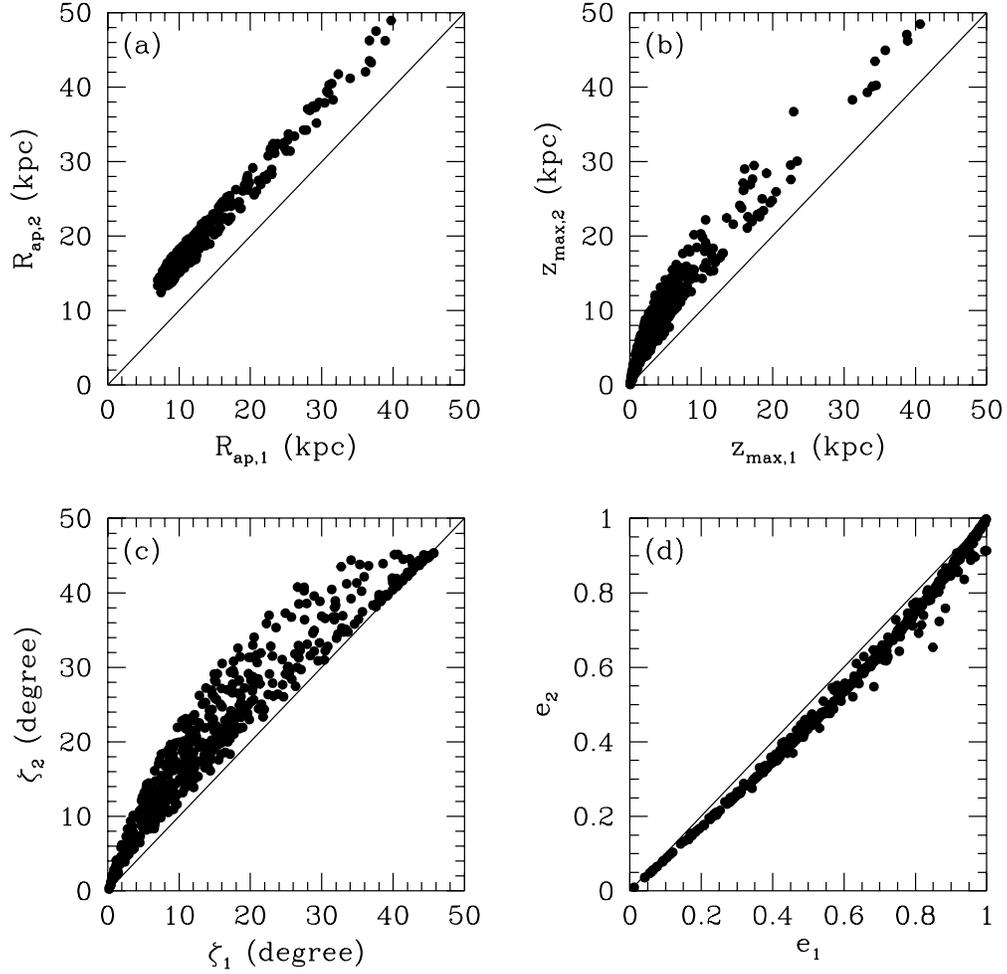}\hfil
\end{center}
\figcaption[fig2.eps]{
Orbital parameters of the sample stars before (ordinate,
subscript 2) and after (abscissa, subscript 1) disk formation, for
(a) apogalactic distance along the plane $R_{ap}$,
(b) maximum height away from the plane $z_{max}$,
(c) inclination angle with respect to the plane
$\zeta = \arctan (z_{max}/R_{ap})$, and
(d) orbital eccentricity $e$.}
\end{figure}

\clearpage
\begin{figure}
\begin{center}
\leavevmode
\epsfxsize=0.7\columnwidth\epsfbox{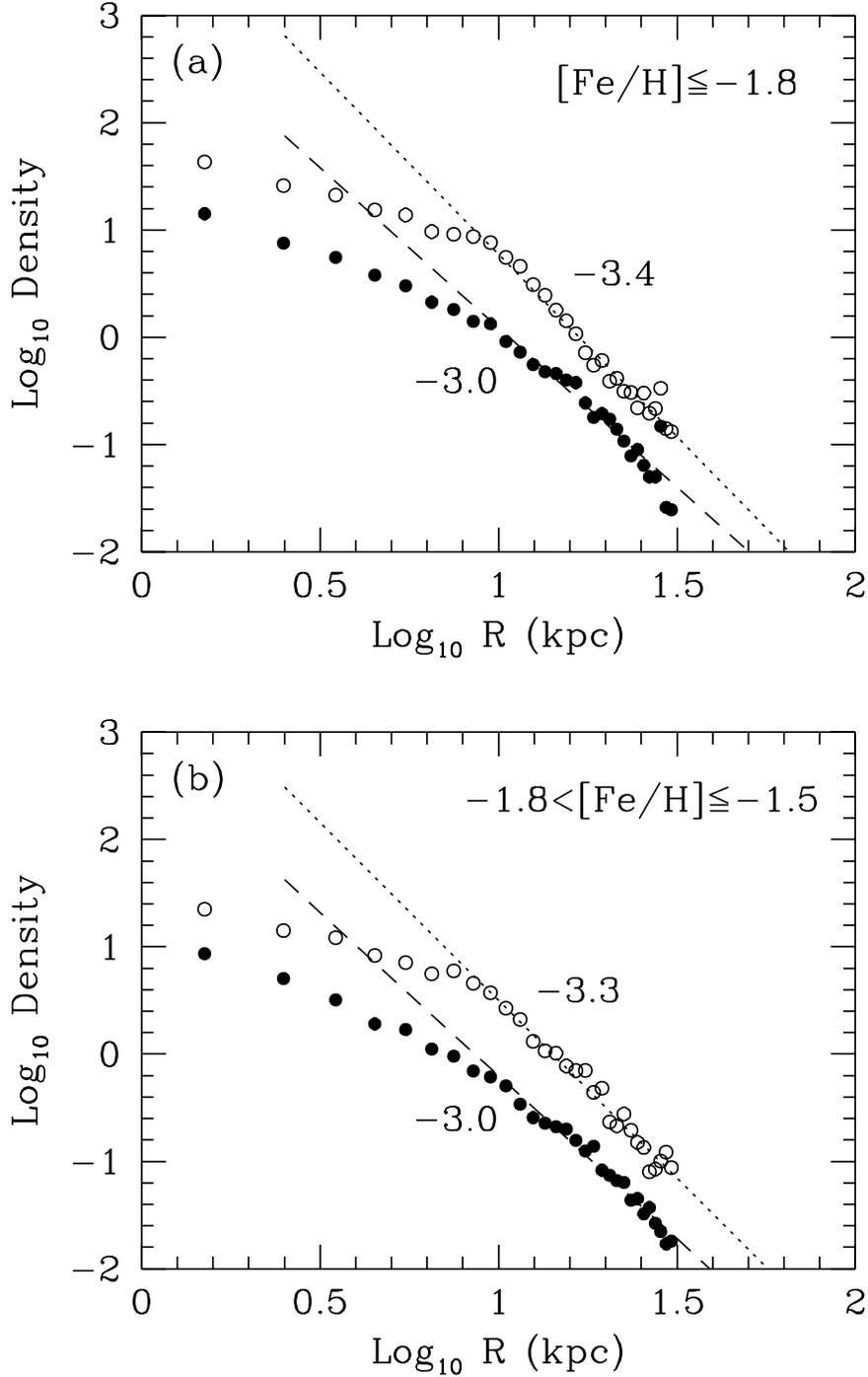}\hfil
\end{center}
\figcaption[fig3.eps]{
Density distributions of the reconstructed halo in
the Galactic plane before (filled circles) and after (open circles) disk
formation, for (a) [Fe/H] $\le-1.8$ and (b) $-1.8<$ [Fe/H]$\le-1.5$.
All plots have been shifted arbitrarily along the vertical axis for clarity.
The dashed and dotted lines denote the best-fit power-law model
with labeled exponents, before and after disk formation, respectively.}
\end{figure}

\clearpage
\begin{figure}
\begin{center}
\leavevmode
\epsfxsize=0.8\columnwidth\epsfbox{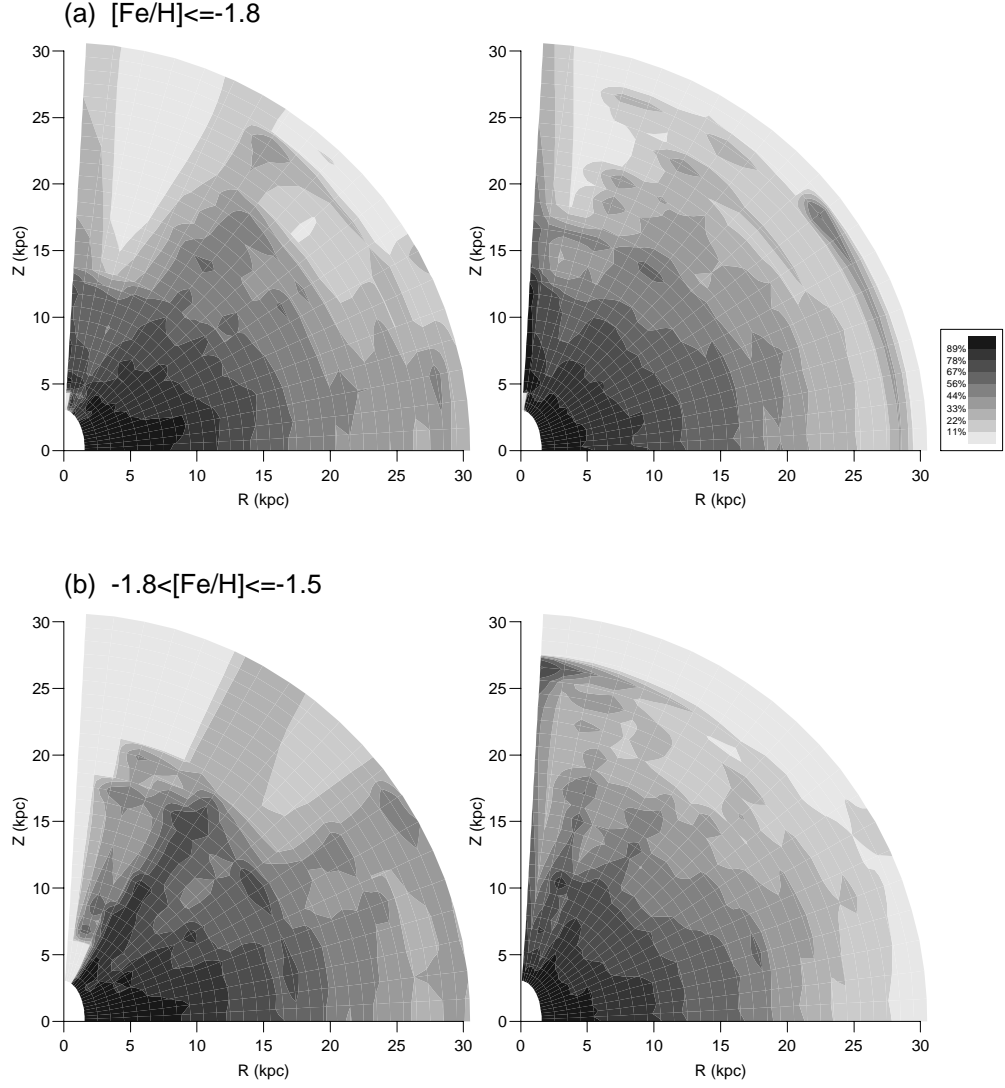}\hfil
\end{center}
\figcaption[fig4.eps]{
Equidensity contours for the reconstructed halo in
the $(R,z)$ plane, for (a) [Fe/H]$\le-1.8$ and (b) $-1.8<$[Fe/H]$\le-1.5$.
Left and right panels correspond to the density distributions after and
before disk formation, respectively.}
\end{figure}

\clearpage
\begin{figure}
\begin{center}
\leavevmode
\epsfxsize=0.7\columnwidth\epsfbox{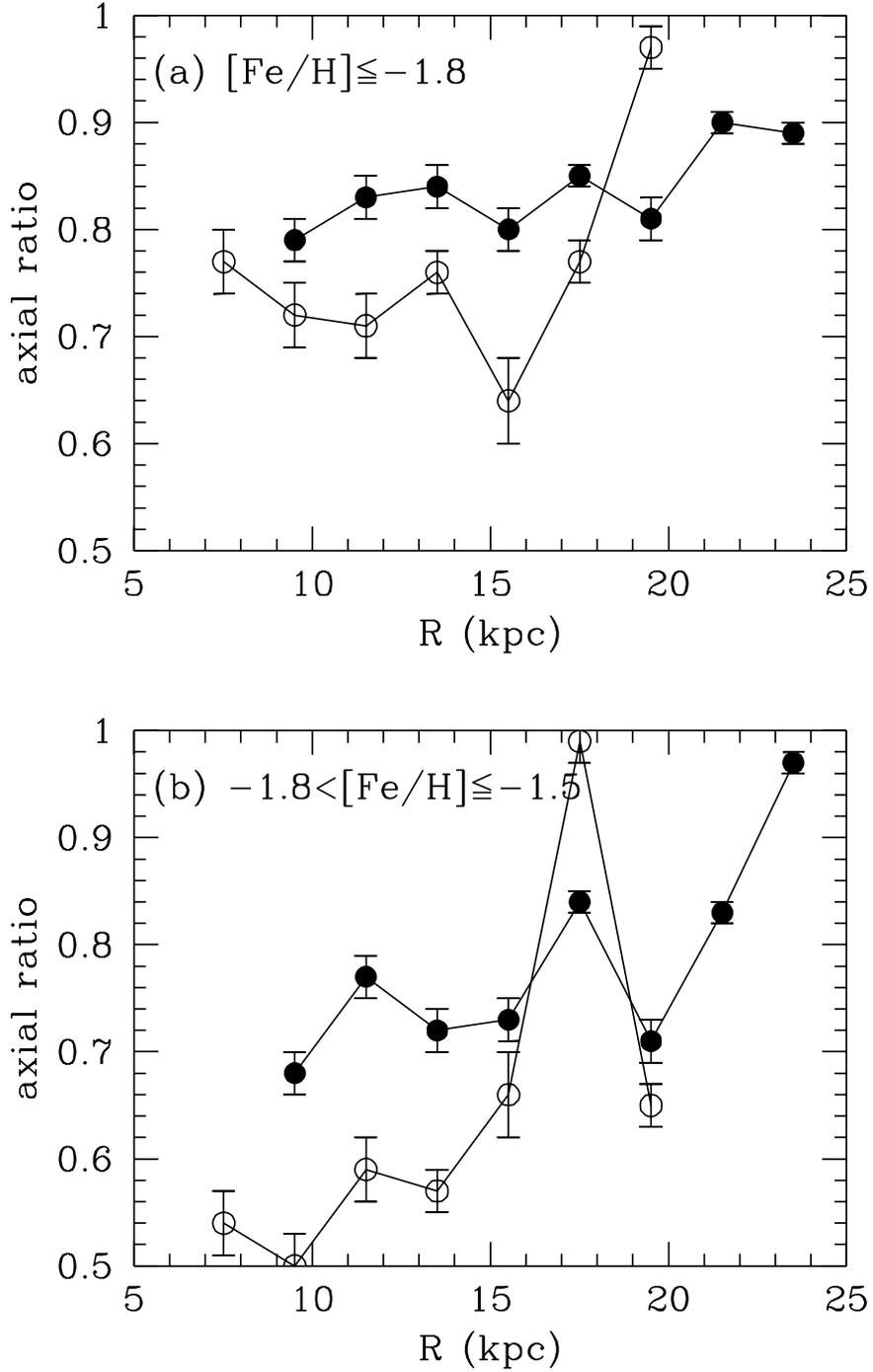}\hfil
\end{center}
\figcaption[fig5.eps]{
Axial ratios for the density distributions of the
reconstructed halo before (filled circles) and after (open circles) disk
formation, for (a) [Fe/H] $\le-1.8$ and (b) $-1.8<$[Fe/H]$\le-1.5$.}
\end{figure}

\clearpage
\begin{figure}
\begin{center}
\leavevmode
\epsfxsize=0.7\columnwidth\epsfbox{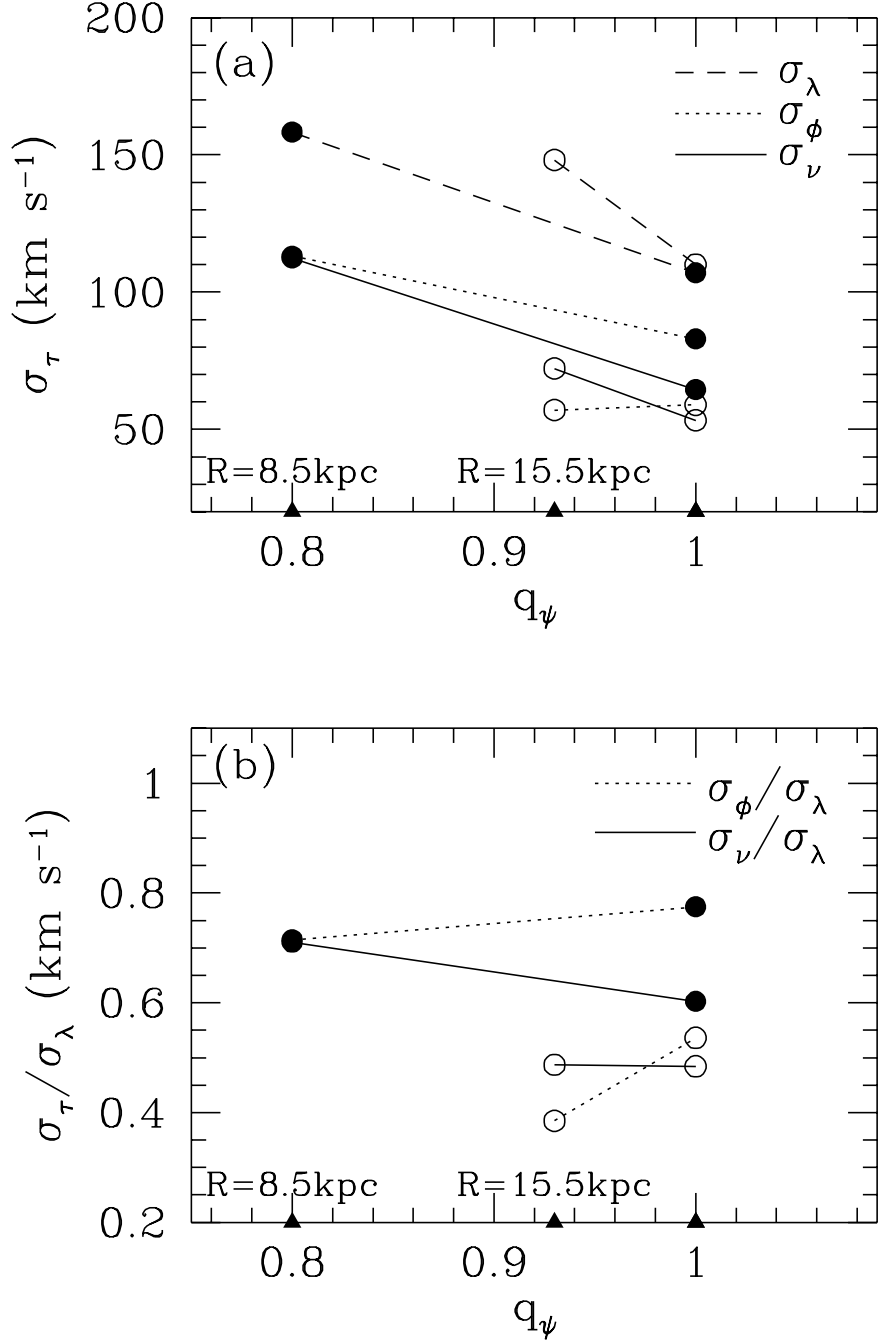}\hfil
\end{center}
\figcaption[fig6.eps]{
(a) Velocity dispersions of the stars with
[Fe/H]$\le-1.8$, $(\sigma_\lambda,\sigma_\phi,\sigma_\nu)$, and
(b) their ratios, $(\sigma_\phi/\sigma_\lambda,\sigma_\nu/\sigma_\lambda)$,
at two radii $R=8.5$ kpc and 15.5 kpc in the Galactic plane ($z=0$).
The abscissa denotes the axial ratio of the potential, $q_\psi$,
at both radii: $q_\psi=0.80$ at $R=8.5$ kpc and 0.93 at $R=15.5$ kpc
when the disk is in place, and $q_\psi=1.00$ at both radii before disk
formation. Left and right hand filled circles show $\sigma_\tau$ or
$\sigma_\tau/\sigma_\lambda$ after and before disk formation, respectively,
for $R=8.5$ kpc, whereas open circles for $R=15.5$ kpc.}
\end{figure}

\end{document}